\documentclass{jltp}

\usepackage{psfig,amstext,amssymb}

\title{Isosbestic points in the spectral function of correlated electrons}

\author{Martin Eckstein, Marcus Kollar, and Dieter Vollhardt}

\address{Theoretical Physics III,
  Center for Electronic Correlations and Magnetism,\\
  Institute for Physics, University of Augsburg,
  D-86135 Augsburg, Germany}

\runninghead{Martin Eckstein, 
  Marcus Kollar, 
  and Dieter Vollhardt}{Isosbestic 
  points in the spectral function of correlated electrons}

\begin{document}

  \maketitle

  \begin{abstract}
    {\rm\footnotesize This paper is dedicated to
      Hilbert v. L\"ohneysen on the occasion of his 60th birthday.}\\
  
    We investigate the properties of the spectral function
    $A(\omega,U)$ of correlated electrons within the Hubbard model and
    dynamical mean-field theory. Curves of $A(\omega,U)$ vs.\ %
    $\omega$ for different values of the interaction $U$ are found to
    intersect near the band-edges of the non-interacting system. For a
    wide range of $U$ the crossing points are located within a sharply
    confined region. The precise location of these ``isosbestic
    points'' depends on details of the non-interacting band structure.
    Isosbestic points of dynamic quantities therefore provide valuable
    insights into microscopic energy scales of correlated systems.

  \end{abstract}

  \section{INTRODUCTION}
  \label{sec:intro}

  A family of non-monotonic curves, obtained by plotting a quantity
  $f(x,y)$ as a function of one of its variables (say, $x$) for
  different values of $y$, will in general intersect. The crossing
  points are located along a curve $x^{*}(y)$ defined by
  \begin{equation}
    \frac{\partial f(x,y)}{\partial y} \bigg|_{x^{*}(y)} =  0\,.
  \end{equation}
  In physics, chemistry and biology these crossing points are
  sometimes found to be confined to a remarkably narrow region, or
  even located at a single point,\cite{Scheibe37} thus leading to a
  conspicuous feature termed \emph{isosbestic
    point}.\cite{Cohen62,isosbestic} In the former case $x^{*}$
  depends only weakly on $y$, while in the latter case $x^{*}$ does
  not depend on $y$ at all.

\begin{figure}
    \centerline{\psfig{clip=true,width=\linewidth,file=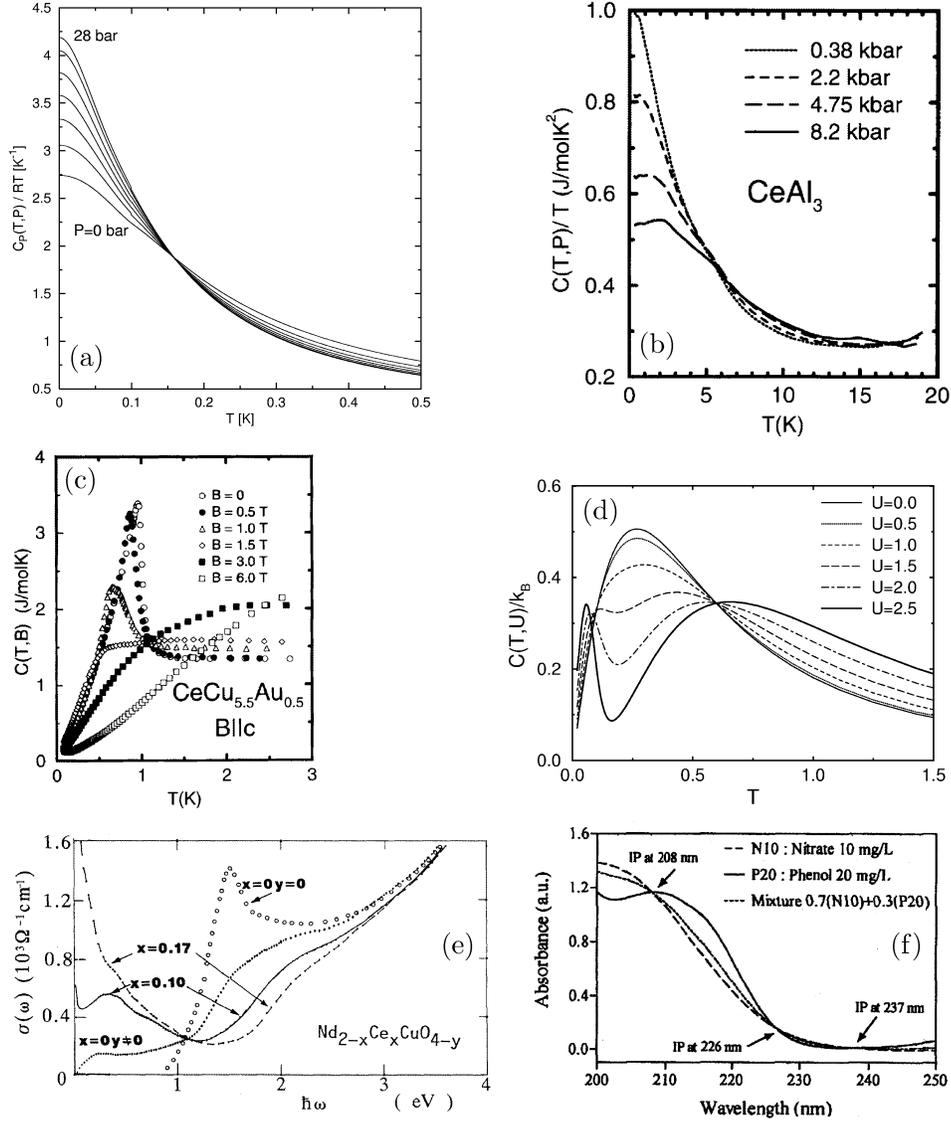}}
    \caption{\label{fig:intro}%
        Examples of crossing points in various correlated systems:
        (a) Specific heat of normal liquid $^3$He
        measured at different pressures\cite{Greywall83}
        (after Ref.~\onlinecite{Vollhardt97}).
        (b) Specific heat of CeAl$_{3}$ measured at different pressures\cite{Brodale86}(after
        Ref.~\onlinecite{Vollhardt97}).
        (c) Specific heat of CeCu$_{5.5}$Au$_{0.5}$
        measured at different magnetic fields\cite{Schlager93}
        (after Ref.~\onlinecite{Vollhardt97}).
        (d) Specific heat of the Hubbard model
        calculated for different interactions\cite{Georges93}
        (after Ref.~\onlinecite{Vollhardt97}).
        (e) Optical conductivity of Nd$_{2-x}$Ce$_{x}$CuO$_{4-y}$
        measured at different doping (after Ref.~\onlinecite{Uchida91}).
        (f) Isosbestic point in ultra\-violet spectra of three solutions
        of phenol and nitrate (after Ref.~\onlinecite{Pouet04}).
     }
  \end{figure}

  For example, the curves of the specific heat $C(T,X)$ vs.\ %
  temperature $T$ of numerous strongly correlated fermionic systems
  are known to cross once or twice when plotted for different values
  of a second thermodynamic variable $X$.\cite{Vollhardt97} In
  particular, crossing points are observed for different pressures
  ($X=P$) in normalfluid $^3$He
  (Fig.~\ref{fig:intro}a)\cite{Greywall83} and heavy-fermion systems
  such as CeAl$_3$ (Fig.~\ref{fig:intro}b).\cite{Brodale86} By
  changing the magnetic field ($X=B$) the same feature is seen in
  heavy-fermion compounds such as CeCu$_{6-x}$Al$_x$
  (Fig.~\ref{fig:intro}c).\cite{Schlager93}

  Crossing points of specific heat curves are also observed in lattice
  models for correlated electrons such as the one-band Hubbard model
  \begin{equation}
    H
    =
    \sum_{ij\sigma}
    t_{ij}\,
    c_{i\sigma}^{\dagger}
    c_{j\sigma}^{\phantom{\dagger}}
    +
    U\sum_{i}n_{i\uparrow}n_{i\downarrow}
    -
    \mu\sum_{i\sigma}n_{i\sigma}
    \,,\label{eq:hubbard}
  \end{equation}
  where $c_{i\sigma}^\dagger$ are creation operators for an electron
  at site $i$ with spin $\sigma$ and $n_{i\sigma}=
  c_{i\sigma}^{\dagger} c_{j\sigma}^{\phantom{\dagger}}$ are density
  operators.  The model contains the hopping amplitude $t_{ij}$ , the
  local Coulomb interaction $U$, and the chemical potential $\mu$ as
  parameters.  At half-filling the curves $C(T,U)$ vs.\ %
  $T$ always cross at two temperatures. This is observed, for example,
  in the case of nearest-neighbor hopping in
  $d=1$,\cite{Shiba72,Juettner98} $d=2$,\cite{Duffy97} and $d=\infty$
  (Fig.~\ref{fig:intro}d),\cite{Georges93} as well as for long-range
  hopping in $d=1$.\cite{Gebhard91}

  As shown in Ref.~\onlinecite{Vollhardt97} the existence of crossing
  points in the specific heat curves, and the fact that they may be
  quite sharp, can be linked (i) to a sum rule for the change of the
  entropy $S(T,X)$ with respect to $X$ in the limit of
  $T\rightarrow\infty$, and (ii) the properties of the
  susceptibilities $\chi^{(n)} (T,X)= \partial^n \xi/ \partial X^n$,
  where $\xi (T,X)$ is the conjugate variable to $X$.  Furthermore,
  the fact that the high-temperature crossing point in $C(T,U)$ for
  the one-band Hubbard model occurs at a nearly universal value
  $C^*/k_{B}\simeq0.34$ was shown to be a consequence of the existence
  of two small parameters: the integral over the deviation of the
  density of states from a constant value, and the inverse dimension,
  $1/d$.\cite{Chandra99}

  The isosbestic points discussed above all appear in curves of the
  specific heat $C(T,X)$ vs.\ %
  temperature $T$ when plotted for different values of another
  thermodynamic variable $X$.  On the other hand, such points are
  known to occur also in dynamic quantities, e.g., in the optical
  conductivity $\sigma(\omega,n)$ of the high-$T_c$ material
  Nd$_{2-x}$Ce$_{x}$CuO$_{4-y}$ where $n$ is the density or doping
  (Fig.~\ref{fig:intro}e),\cite{Uchida91} and in the Raman response
  $\chi(\omega, T)$ of the Hubbard model.\cite{Freericks} In general,
  there is no reason for the intersection of these curves to occur at
  one sharp frequency $\omega^*$, i.e., to be completely independent
  of any other parameter defining the family of curves.  There are,
  however, at least two classes of isosbestic points which are
  genuinely point-like. One is the exact crossing of curves described
  by a scaling function in critical phenomena; see, for example, the
  crossing of conductance curves near the Anderson
  transition.\cite{Ohtsuki} Another one is found in optical
  spectroscopic studies\cite{Scheibe37,Cohen62,Pouet04} of systems
  consisting of two components with densities $n_a$, $n_b$ and
  $n=n_a+n_b=\text{const}$. In this case the absorbance as a function
  of frequency $\alpha(\omega,n_{a})$ depends only linearly on the
  density, i.e., has the special form
  \begin{equation}
    \alpha(\omega,n_{a})=n_{a}f_{a}(\omega)+(n-n_{a})f_{b}(\omega).
  \end{equation}
  If $f_{a}$ and $f_{b}$ coincide at a frequency $\omega^*$, i.e.,
  $f_{a}(\omega^*)=f_{b}(\omega^*)$, then
  \begin{equation}
    \frac{\partial \alpha(\omega,n_{a})}{\partial n_{a}} \bigg|_{\omega^{*}} =
    0\,.
  \end{equation}
  This implies that for all densities $n_{a}$ the absorbance curves
  intersect at one frequency $\omega^*$ (or the equivalent wave
  length, Fig.~\ref{fig:intro}f).  Similar isosbestic points are found
  in diffraction experiments on glasses.\cite{Johnson85} Quite
  generally, whenever a system is a superposition of two (or more)
  components such that its dynamic quantities, e.g., the dynamic
  conductivity or a response function, have the form described by
  Eq.~(3), isosbestic points are bound to occur. This applies in
  particular to any kind of two-fluid model employed, for example, in
  phenomenological theories of superconductivity and superfluidity.
  There the density of the two components (e.g., the normal and
  superfluid component) depend on temperature while the total density
  is constant: $n=n_{a}(T)+n_{b}(T)=\text{const}$. Properties of the
  system are then described by the superposition of the two
  components, leading to a special dependence of quantities $f(T,X)$
  on $T$ and $X$ of the form
  \begin{equation}
    f(T,X)=n_{a}(T)f_{a}(X)+[n-n_{a}(T)]f_{b}(X)\,.
  \end{equation}
  This implies the crossing of curves for different temperatures $T$
  at a single point $X^*$ determined by $f_{a}(X^*)=f_{b}(X^*)$.
  Whether the surprisingly sharp isosbestic points found, for example,
  in the optical conductivity of Nd$_{2-x}$Ce$_{x}$CuO$_{4-y}$
  (Fig.~\ref{fig:intro}e)\cite{Uchida91} can be explained in this way
  still has to be investigated.

  Disregarding critical phenomena with scaling behavior and the
  special linear dependence given by Eq.~(3), isosbestic points in
  \emph{dynamical} quantities of correlated electron systems have so
  far only been noticed but never explained. As in the case of the
  crossing of specific heat curves there are two separate questions to
  be answered: (i) Why do curves of frequency-dependent quantities
  cross at all, and (ii) under what circumstances is the crossing
  region confined to a narrow region, or is even point-like?  In
  analogy to the entropy sum rule in the case of specific heat curves,
  a good starting point for such an investigation is the study of
  \emph{frequency sum rules}.  Examples are the f-sum rule for the
  dynamical conductivity and sum rules involving the spectral
  function.

  In this paper we investigate crossing points in frequency by
  studying a particularly basic quantity --- the local
  (momentum-integrated) spectral function $A(\omega,U)$ of the Hubbard
  model, a quantity which obeys fundamental frequency sum rules.  In
  the following we will write $A(\omega,U) \equiv A(\omega)$; the
  parameter $U$ will only be written when explicitly needed.  For
  simplicity we will employ the Bethe lattice in the limit of infinite
  coordination number. In this case the self-energy $\Sigma$ becomes
  local and can be calculated using dynamical mean-field theory
  (DMFT).\cite{Metzner89,Georges92,Vollhardt93,Jarrell95,Georges96,PT}
  In Sec.~\ref{subsec:specfunc} we discuss the behavior of $A(\omega)$
  and in Sec.~\ref{subsec:xings} the location of crossing points.
  Their behavior at small interaction is studied in
  Sec.~\ref{sec:weak}.  We use weak-coupling perturbation theory
  (Sec.~\ref{subsec:wct}) to investigate how the crossing points of
  $A(\omega)$ depend on the non-interacting bandstructure and the
  interaction strength $U$. This is performed for both the Bethe
  lattice (Sec.~\ref{subsec:wcb}) and a model density of states
  (Sec.~\ref{subsec:wcm}), which differ in their van-Hove
  singularities.  A conclusion in Sec.~\ref{sec:concl} closes the
  paper.

  \section{CROSSING POINTS OF THE SPECTRAL FUNCTION}
  \label{sec:xings}

  \subsection{Spectral function in DMFT}
  \label{subsec:specfunc}

\begin{figure}[t]
    \centerline{\psfig{clip=true,width=0.85\linewidth,file=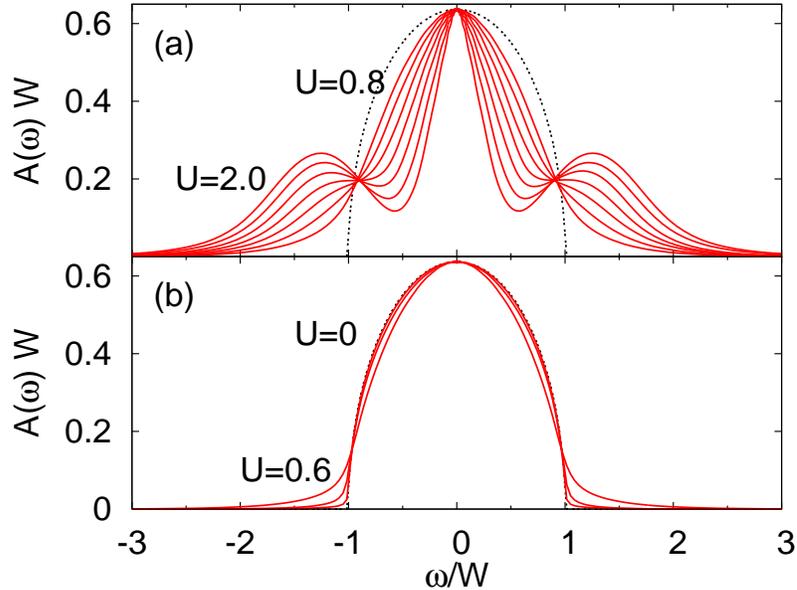}}
    \caption{Spectral function of the Hubbard model, calculated with
      DMFT for the Bethe DOS [Eq.~(\ref{eq:rho-bethe})]; the data are
      from Ref.~\onlinecite{Bulla98}, (a) Sharp crossing points
      (isosbestic points) occur for intermediate $U$ ($U/W =
      0.8,\,1.0,\ldots,2.0$).  (b) For small $U$ the crossing points
      move up and inward as $U$ is increased ($U/W =
      0.2,\,0.4,\,0.6$).}
    \label{fig:nrg-g}
  \end{figure}

  We consider the paramagnetic phase of the Hubbard model
  (\ref{eq:hubbard}) in
  DMFT.\cite{Metzner89,Georges92,Vollhardt93,Jarrell95,Georges96,PT}
  In this case $\Sigma$ does not have any site- or spin-dependence,
  and the local Green function $G(\omega)$ becomes
  \begin{equation}
    G(\omega) = G_0(\omega+\mu-\Sigma(\omega))\,.
    \label{eq:G}
  \end{equation}
  The non-interacting Green function $G_0(z)$, written as a function of
  the complex variable $z$, is given by the Hilbert transform of the
  non-interacting density of states (DOS) $\rho(\epsilon)$,
  \begin{equation}
    G_0(z)=\int
    d\epsilon\frac{\rho(\epsilon)}{z-\epsilon}\,.
    \label{eq:G0}
  \end{equation}
  The local spectral function $A(\omega)$ is then given by
  \begin{equation}
    A(\omega)
    =
    -\frac{1}{\pi}
    G''(\omega+i0)
    \,.
    \label{eq:a}
  \end{equation}
  (We use single and double primes to indicate real and imaginary part
  of a complex quantity.)  In Fig.~\ref{fig:nrg-g} the spectral
  function is plotted for various $U$ in the case of half-filling.
  The numerical data are taken from Ref.~\onlinecite{Bulla98}, where
  the effective impurity problem was solved with numerical
  renormalization group (NRG).  As usual for model calculations within
  DMFT the semi-elliptic Bethe DOS
  \begin{equation}
    \rho(\epsilon)=\frac{2}{\pi W }\sqrt{1-\frac{\epsilon^2}{W^2}},
    \quad W=1,
    \label{eq:rho-bethe}
  \end{equation}
  with half-bandwith $W$ was used as non-interacting DOS; we set $W=1$
  as energy scale.

  For all values of $U$ in Fig.~\ref{fig:nrg-g} the system is assumed
  to be in the Fermi-liquid phase. As the interaction increases,
  spectral weight is redistributed from the vicinity of the Fermi
  level to the Hubbard subbands which are peaked at frequency
  $\omega\approx\pm U/2$, and the well-known three-peak structure of
  the spectral density emerges.  The central spectral peak
  vanishes at the metal-insulator transition, which occurs at
  $U=2.92\,W$.\cite{Bulla01,BluemerDiss}

  \subsection{Crossing points}
  \label{subsec:xings}

  There are three crossing points visible in Fig.~\ref{fig:nrg-g}.
  The first one, at $\omega=0$, has a simple explanation: In infinite
  dimensions Luttinger's theorem entails that the chemical potential
  $\mu$ in the interacting system must be shifted according
  to\cite{Muellerhartmann89b}
  \begin{equation}
    \mu = \mu_0 + \Sigma(0)
    \label{eq:luttinger}
  \end{equation}
  in order to keep the number of particles fixed, where $\mu_0$ is the
  Fermi energy for the non-interacting system with the same number of
  particles.  Thus the value of the Green function
  $G(\omega)=G_0(\omega+\mu-\Sigma(\omega))$ at the Fermi level is
  independent of $U$ in the metallic phase, and the crossing point at
  $\omega=0$ is exact.  That the curves do not really cross but only
  touch is a consequence of the particle-hole symmetry at
  half-filling, which implies $A(\omega)=A(-\omega)$.  For any other
  density or an asymmetric DOS there is a true crossing point at
  $\omega=0$.

  The other two crossing points of $A(\omega)$ are situated at $\omega
  \approx \pm 1$.  In contrast to the pinning at $\omega=0$, their
  explanation is more complicated.  The crossing region at
  $\omega\approx \pm1$ is very narrow for intermediate values of $U$
  ($0.8 \lesssim U/W \lesssim 2.0$, Fig.~\ref{fig:nrg-g}a).  For
  smaller values of $U$ (Fig.~\ref{fig:nrg-g}b), we observe that the
  region of crossing points is less confined.  The crossing points
  move up and inward as $U$ is increased.  However, as shown below the
  crossing \emph{frequency} does not depend strongly on $U$ up to
  $U\approx 2$.

  Let us first establish that curves $A(\omega,U)$ vs.\ %
  $\omega$ for different values of $U$ always cross at some frequency
  $\omega^*$, defined by
  \begin{equation}
    \left.
      \frac{\partial A(\omega,U)}{\partial U}
    \right|_{\omega^*(U)}
    =0.
    \label{eq:a-crossing}
  \end{equation}
  That two such curves \emph{must} cross follows from the sum rules
  \begin{eqnarray}
    \int_{-\infty}^0 d\omega A(\omega,U) & = & \frac{n}{2}, \\
    \int_{-\infty}^{\infty} d\omega A(\omega,U) & = & 1,
    \label{eq:sum-rule}
  \end{eqnarray}
  where $n$ is the density.  The derivatives of these equations with
  respect to $U$ yields
  \begin{equation}
    \int_{-\infty}^0 d\omega
    \frac{\partial A(\omega,U)}{\partial U}
    =
    \int_{0}^{\infty} d\omega
    \frac{\partial A(\omega,U)}{\partial U}
    =0\,,
  \end{equation}
  and thus $\partial A(\omega,U) /\partial U$ vanishes either
  identically or changes sign at least at one point in each of the
  intervals $(-\infty,0)$ and $(0,\infty)$.  This is the location of
  the crossing point.

\begin{figure}
    \centerline{\psfig{clip=true,width=0.85\linewidth,file=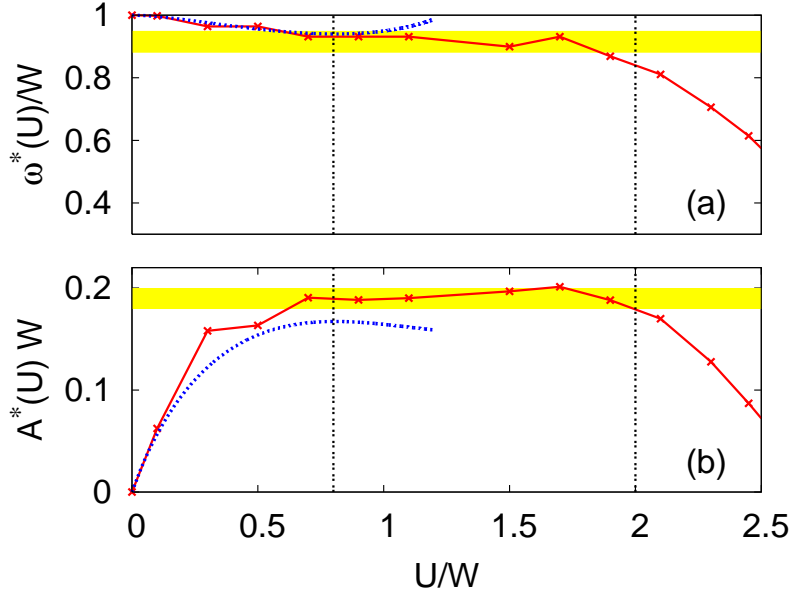}}
    \caption{(a) The crossing point $\omega^*(U)$ of the spectral
      function for the half-filled Hubbard model as a function of $U$
      and (b) $A^*(U)$, the value of $A(\omega)$ at this point.  Each
      data point corresponds to the intersection of two neighboring
      curves shown in Fig.~\ref{fig:nrg-g}.  The dotted lines delimit
      the range of $U$ in Fig.~\ref{fig:nrg-g}a and b, and the shaded
      bars mark the corresponding width of the crossing region
      Fig.~\ref{fig:nrg-g}a.  Dotted curves are the result of
      second-order perturbation theory (Sec.~\ref{sec:weak}).  Due to
      particle-hole symmetry of the half-filled system it is
      sufficient to consider only $\omega>0$.}
    \label{fig:nrg-crossing}
  \end{figure}

  However, the existence of a solution to Eq.~(\ref{eq:a-crossing}) is
  not sufficient for the observation of a sharp crossing point as in
  Fig.~\ref{fig:nrg-g}a.  In general the crossing frequency
  $\omega^*$, and also the value $A^*\equiv A(\omega^*(U),U))$ of
  $A(\omega)$ at this point depends on $U$.  This becomes directly
  evident from Fig.~\ref{fig:nrg-crossing}, where $\omega^*(U)$ and
  $A^*(U)$ are plotted as a function of $U$.  Since these data were
  obtained from the intersections of curves $A(\omega,U)$ differing by
  $\Delta U = 0.1\,W$, while a solution to Eq.~(\ref{eq:a-crossing})
  corresponds to taking $\Delta U\to0$, they have to be understood as
  rather rough estimates for $\omega^*$ and $A^*$.  Furthermore, the
  resolution of the NRG data at $\omega=\pm W$ is already much lower
  than in the vicinity of the Fermi level.  Nevertheless,
  Fig.~\ref{fig:nrg-crossing} clearly shows that the strong variation
  of $A^*$ with $U$ (for small $U$) causes the crossing region to
  become much narrower when only intermediate $U$ are plotted as in
  Fig.~\ref{fig:nrg-g}a.  By contrast the value of $\omega^*$ remains
  close to the non-interacting band-edge $\omega=W$ for all $U$ up to
  $U \approx 2\,W$.

  The variation of $A^*$ and $\omega^*$ with $U$, i.e., the
  derivatives $d\omega^{*}/dU$ and $dA^*/dU$ are not independent. The
  latter is given by
  \begin{equation}
    \frac{d  A^*}{dU} =
    \left. \frac{\partial A(\omega,U)}
      {\partial U}\right|_{\omega^*}+
    \left. \frac{\partial A(\omega,U)}
      {\partial \omega}\right|_{\omega^*}\frac{d\omega^*}{dU}\,,
  \end{equation}
  and since the first term vanishes by definition
  [Eq.~(\ref{eq:a-crossing})] we obtain the basic relation
  \begin{equation}
    \frac{d  A^*}{dU} =
    \left. \frac{\partial A(\omega,U)}
      {\partial \omega}\right|_{\omega^*}\frac{d\omega^*}{dU}.
  \end{equation}
  In the present case $d\omega^{*}/dU$ is small for all $U<2W$, while
  the derivative $\partial A /\partial \omega|_{\omega^*}$ diverges as
  $U$ approaches zero. This is because $A(\omega)$ approaches the
  non-interacting DOS $\rho(\epsilon)$ [Eq.~(\ref{eq:rho-bethe})] and
  $\omega^*$ moves to the band-edge where $\rho(\epsilon)$ has a
  van-Hove singularity.  Thus an almost constant crossing frequency
  $\omega^*$ is not in contradiction to a large variation of $A^*(U)$.

  \section{CROSSING POINTS AT WEAK COUPLING}
  \label{sec:weak}

  As discussed above, well-defined crossing points in $A(\omega)$
  exist not only at intermediate $U$ (Fig.~\ref{fig:nrg-g}a) but also
  at small $U$ (Fig.~\ref{fig:nrg-g}b).  We now investigate the
  behavior of $\omega^*$ in the limit of small $U$ by means of
  many-body weak-coupling perturbation theory (Sec.~\ref{subsec:wct}).
  We first show that $\omega^*$ approaches the band-edge as $\omega^*
  = 1 + \mathcal{O}(U^2)$ if there is a van-Hove singularity at the
  band-edge (i.e., a divergent slope of $\rho(\epsilon)$, see
  Sec.~\ref{subsec:wcb}), and later contrast this by considering a DOS
  which vanishes linearly at the band-edge (Sec.~\ref{subsec:wcm}).

  \subsection{Weak-coupling perturbation theory}
  \label{subsec:wct}

  We start with a short summary of weak-coupling perturbation theory
  for the local Green function in the paramagnetic phase of the
  Hubbard model. For $U\ll W$ one can use a perturbative expansion
  \begin{equation}
    \Sigma(\omega) = \sum_{n\ge 1} U^n \Sigma_n (\omega)
    \label{eq:sigma-expansion}
  \end{equation}
  for the self-energy $\Sigma$, where the functions $\Sigma_n(\omega)$
  are given by the sum over all irreducible Feynman diagrams with $n$
  vertices.  It is convenient to perform this expansion relative to
  the Hartree approximation.\cite{Schweitzer91} The series
  (\ref{eq:sigma-expansion}) is then rearranged such that each line in
  the Feynman diagrams for $\Sigma_n$ is replaced by the Hartree
  expression for the Green function, and in turn all Feynman diagrams
  which contain first-order self-energy insertions are omitted.  Then
  $\Sigma_1$ is just given by the static Hartree self-energy
  $\Sigma_{\text{H}} = Un/2$, and only a single Feynman diagram
  contributes to $\Sigma_2(\omega)$.

  In the following we discuss only the case $\mu=U/2$, for which the
  Hubbard model on a bipartite lattice becomes half-filled and
  particle-hole symmetric.  This entails a number of important
  simplifications:\cite{Gebhardt03} All odd terms in
  Eq.~(\ref{eq:sigma-expansion}) apart from $\Sigma_{\text{H}}$ vanish
  and for the even terms the symmetry relations $\Sigma_n'(\omega) =
  -\Sigma_n'(-\omega)$ and $\Sigma_n''(\omega)=\Sigma_n''(-\omega)$
  hold, so that $\Sigma(0)=\Sigma_{\text{H}}=U/2$. This is of course
  consistent with Eq.~(\ref{eq:luttinger}).

  The calculation of diagrams is considerably simplified in infinite
  dimensions, where multidimensional momentum integrals can be reduced
  to one-dimensional integrals over the non-interacting density of
  states.  For further reference we mention a convenient expression
  for the second-order diagram. In case of particle-hole symmetry
  $\Sigma_2(\omega)$ can be written as\cite{Schweitzer91}
  \begin{equation}
    \Sigma_2(\omega) =
    ~\raisebox{-1.2mm}{\psfig{height=4mm,file=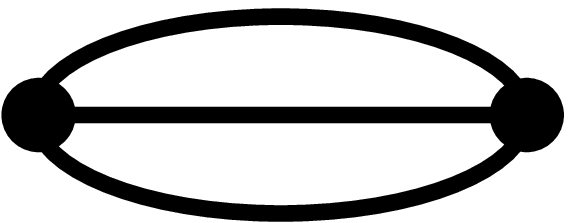}}~
    =
    -i \int_{-\infty}^{\infty}
    d\lambda\,e^{i\lambda z}
    [2A'(\lambda)^3-6A''(\lambda)^2A'(\lambda)],
    \label{eq:sigma2}
  \end{equation}
  where
  \begin{equation}
    A(\lambda)
    =
    \int_{-\infty}^{\infty} d\xi\,e^{-i\lambda\xi}\rho(\xi) f(\xi)
    \label{eq:a_lambda}
  \end{equation}
  and $f(\xi)=1/(e^{\beta\xi}+1)$ is the Fermi function.

  \subsection{Bethe density of states}
  \label{subsec:wcb}

  As a representative of a DOS with a van-Hove singularity at the
  band-edge we first study the Bethe DOS [Eq.~(\ref{eq:rho-bethe})].
  To solve Eq.~(\ref{eq:a-crossing}) for $\omega^*$ we insert
  Eqs.~(\ref{eq:a}) and (\ref{eq:G}) and obtain
  \begin{equation}
    \frac{\partial A}{\partial U} \propto
    \left[\frac{d G_0(z)}{dz}
      \left(
        \frac{\partial \mu}{\partial U}-
        \frac{\partial \Sigma( \omega^*)}{\partial U}
      \right)
    \right]''
    _{z=\omega^*+\mu-\Sigma(\omega^*)}
    =0
    \,.
    \label{eq:a-crossing2}
  \end{equation}
  Here the chemical potential $\mu$ has to be differentiated with
  respect to $U$ because the spectrum in (\ref{eq:a-crossing})
  corresponds to a given density.  One can use
  Eq.~(\ref{eq:luttinger}) to rewrite this derivative as $\partial
  \mu/\partial U=\partial \Sigma(0)/\partial U$.  From
  Eq.~(\ref{eq:luttinger}) and the fact that the first-order
  self-energy $\Sigma_{\text{H}}$ is static, it follows that
  $\mu-\Sigma(\omega)=\mathcal{O}(U^2)$ for $U\to0$.  Thus $\partial
  \mu/\partial U-\partial\Sigma/\partial U$ vanishes identically. In
  particular one has $\mu-\Sigma(\omega) = -U^2 \Sigma_2(\omega) +
  \mathcal{O}(U^4)$ for half-filling. However, instead of looking for
  a sign change in $\partial A/\partial U$ as in
  Eq.~(\ref{eq:a-crossing}) one can more conveniently investigate the
  sign changes in $\partial A/\partial (U^2)$.
  Eq.~(\ref{eq:a-crossing2}) for the crossing frequency at $U=0$,
  \begin{equation}
    \omega_0 \equiv
    \lim_{U\to0} \omega^*(U)\,,
  \end{equation}
  then becomes
  \begin{equation}
    \frac{\partial A}{\partial (U^2)} \propto
    \Sigma_2''(\omega^*)
    \left(\frac{d G_0}{dz}\right)'_{z=\omega^*}
    +
    \Sigma_2'(\omega^*)
    \left(\frac{d G_0}{dz}\right)''_{z=\omega^*}
    =0.
    \label{eq:a-crossing3}
  \end{equation}

  For the Bethe DOS the non-interacting Green function (\ref{eq:G0})
  and its derivative $dG_0/dz$ are given in analytical form as
  \begin{equation}
    G_0(z)=2(z-\sqrt{z-1}\sqrt{z+1}),
    \qquad
    \frac{dG_0}{dz}=2-\frac{2z}{\sqrt{z-1}\sqrt{z+1}}
    \,,
    \label{eq:g-bethe}
  \end{equation}
  where the complex square root denotes the principal branch.  When
  solving Eq.~(\ref{eq:a-crossing2}) for small $U$, special care has
  to be taken to handle the singularity of $dG_0/dz$ at the band-edges
  correctly. The leading contribution is given by
  \begin{equation}
    \left.\frac{dG_0}{dz}\right|_{z=\omega+i0}
    \sim
    \left\{
      \begin{array}{cc}
        i \frac{\sqrt{2}}{\sqrt{1-\omega}}
        &
        \text{for}
        \quad
        \omega \uparrow 1
        \\
        \frac{-\sqrt{2}}{\sqrt{\omega-1}}
        &
        \text{for}
        \quad
        \omega \downarrow 1
      \end{array}
    \right.\,.
  \end{equation}
  The perturbative self-energy $\Sigma_2(\omega)$ is continuous and at
  least differentiable once as seen from Eq.~(\ref{eq:sigma2}) and
  (\ref{eq:a_lambda}).  To obtain $\partial A/\partial(U^2)$ close to
  $\omega=1$ we can thus replace $\Sigma_2$ by its value at the
  band-edge, $\Sigma''_2(1)\equiv v_2 \approx -0.2926$ and
  $\Sigma'_2(1)\equiv u_2 \approx -0.0692$.  The result is
  \begin{equation}
    \frac{\partial A(\omega)}{\partial U^2}
    \sim
    \left\{
      \begin{array}{cc}
        \frac{\sqrt{2}}{\sqrt{1-\omega}}\,u_2
        &
        \text{for}
        \quad
        \omega \uparrow 1
        \\
        \frac{-\sqrt{2}}{\sqrt{\omega-1}}\,v_2
        &
        \text{for}
        \quad
        \omega \downarrow 1
      \end{array}
    \right.
    \label{eq:asym-a}
  \end{equation}
  for $U\to0$. Because $v_2/u_2>0$ this implies a sign change of the
  derivative $\partial A/\partial (U^2)$ at $\omega=1$, and thus the
  crossing point tends towards the band-edge in the limit of small
  $U$.

  Because the self-energy has an expansion only in even powers of $U$
  one might expect that the same is true also for the crossing point
  $\omega^*(U)$ as a function of $U$.  In particular, this would imply
  that for small $U$ the crossing frequency varies only weakly in the
  sense that the linear term $\omega^*\sim U$ is absent.  On the other
  hand, the square root in $G_0(z)$ at $z=1$ may cause $\omega^*(U)$
  to become nonanalytic at $U=0$.  However, as we will now show that
  these nonanalytic contributions occur only in higher order terms,
  and that the solution of (\ref{eq:a-crossing2}) for small $U$ is of
  the type
  \begin{equation}
    \omega^*= 1 + a U^2 + \mathcal{O}(|U|^3)\,.
    \label{eq:o-expansion}
  \end{equation}
  To single out the nonanalyticity of $dG_0/dz$ in
  Eq.~(\ref{eq:a-crossing2}) we introduce the notation
  $\omega^*=1+\delta\omega$ and $z=\omega^*-\Sigma(\omega^*) \equiv
  1+\delta z$, i.e., $\delta z = \delta \omega - \Sigma(\omega^*)$.
  Eq.~(\ref{eq:a-crossing2}) then reads
  \begin{equation}
    1 =
    \left(\delta z^{-\frac{1}{2}}\right)'' C_1
    +
    \left(\delta z ^{-\frac{1}{2}}\right)' C_2,
  \end{equation}
  where we introduced
  \begin{eqnarray}
    C_{1} &=&
    \left(
      \frac{1+\delta z}{\sqrt{2+\delta z}}
      \frac{\partial (\Sigma-\Sigma_{\text{H}})}{\partial U}
    \right)'
    \Big/
    \left(
      \frac{\partial (\Sigma-\Sigma_{\text{H}})}{\partial U}
    \right)'',
    \\
    C_{2} &=&
    \left(
      \frac{1+\delta z}{\sqrt{2+\delta z}}
      \frac{\partial (\Sigma-\Sigma_{\text{H}})}{\partial U}
    \right)''
    \Big/
    \left(
      \frac{\partial (\Sigma-\Sigma_{\text{H}})}{\partial U}
    \right)''.
  \end{eqnarray}
  Let us now write $\delta z \equiv r e^{i\phi}$ whence we obtain
  \begin{equation}
    \sqrt{r} =
    C_2 \cos \frac{\phi}{2}
    -C_1 \sin \frac{\phi}{2}.
    \label{eq:r=cos-sin}
  \end{equation}
  With the ansatz (\ref{eq:o-expansion}) one has
  $\sqrt{r}=\mathcal{O}(|U|)$ and $\tan\phi= \delta z '' / \delta z '
  $ $\sim$ $ -v_2/(a-u_2) + \mathcal{O}(|U|)$ as $U\to 0$, whereas
  $C_1 = v_2/u_2 + \mathcal{O}(U^2) $ and $C_2 = 1+ \mathcal{O}(U^2)$.
  Thus (\ref{eq:o-expansion}) solves Eq.~(\ref{eq:a-crossing2})
  provided that
  \begin{equation}
    v_2 \cos \frac{\phi}{2}
    -u_2\sin \frac{\phi}{2}
    \to 0
  \end{equation}
  for $U\to 0$. The solution of this, $\tan(\phi/2) \to v_2/u_2$,
  together with the behavior of $\tan\phi$ noted above, yields after
  some manipulation
  \begin{equation}
    a =  \frac{u_2}{2}\left(1+\frac{v_2^2}{u_2^2}\right) = 0.653.
  \end{equation}

  The results of this section are easily generalized for any
  non-interacting DOS with a van-Hove singularity $\rho(\omega) \sim
  \sqrt{|\omega-W|}$ at the band-edge $\omega=\pm W$, as, e.g., for
  the DOS for a simple cubic lattice.  This is because the dependence
  of $G_0(z)=\int d\epsilon\, \rho(\epsilon)/(z-\epsilon)$ and
  $dG_0/dz = -\int d\epsilon\,\rho(\epsilon)/(z-\epsilon)^2$ on
  $\rho(\epsilon)$ is linear. The singular part of $dG_0(z)/dz$ can be
  split off and the remaining continuous part does not enter the
  first-order result for $\omega^*$.  We thus conclude that a van-Hove
  singularity at a band-edge implies that in its vicinity a crossing
  point in the spectrum exists for small $U$, provided that
  $\Sigma_2'(1)$ and $\Sigma_2''(1)$ have the same sign.

  \subsection{Model density of states}
  \label{subsec:wcm}

  The situation is different if there is no van-Hove singularity at
  the band-edge of the non-interacting DOS, e.g., if the DOS vanishes
  linearly. In this case its derivative at the band-edge does not
  diverge. Then the weak-coupling crossing point $\omega_0$ need not
  be located at the band-edge.  We now demonstrate this by
  investigating the family of model functions
  \begin{equation}
    \rho_{\alpha}(\epsilon)=
    \frac{\alpha+1}{2\alpha W}(1-|\epsilon/W|^{\alpha})
    \Theta(W-|\epsilon|),
    \quad W=1,
    \alpha=1,2,3,\ldots
    \label{eq:model-dos}
  \end{equation}
  as non-interacting DOS.  Here $\Theta(x)$ is the step function and
  again the half band-width $W=1$ has been chosen as energy unit.
  \begin{figure}
    \centerline{%
      \psfig{clip=true,height=5.9cm,file=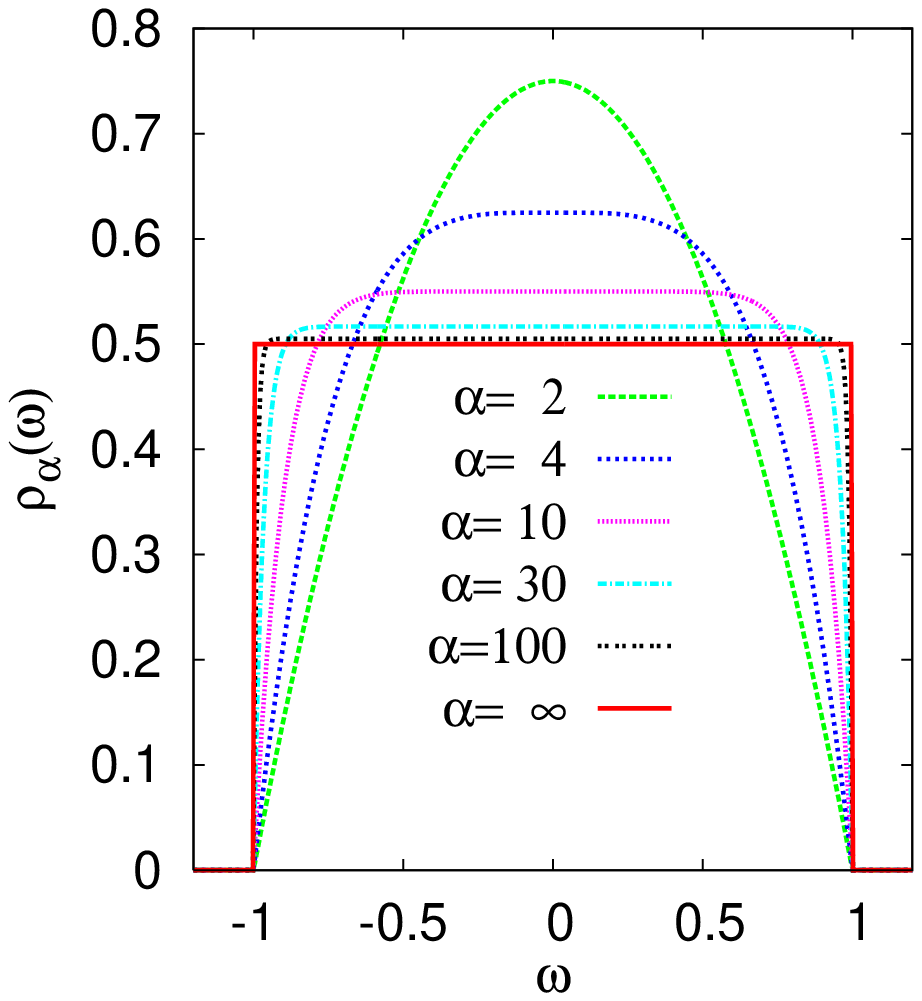}%
      \hspace*{\fill}%
      \psfig{clip=true,height=5.9cm,file=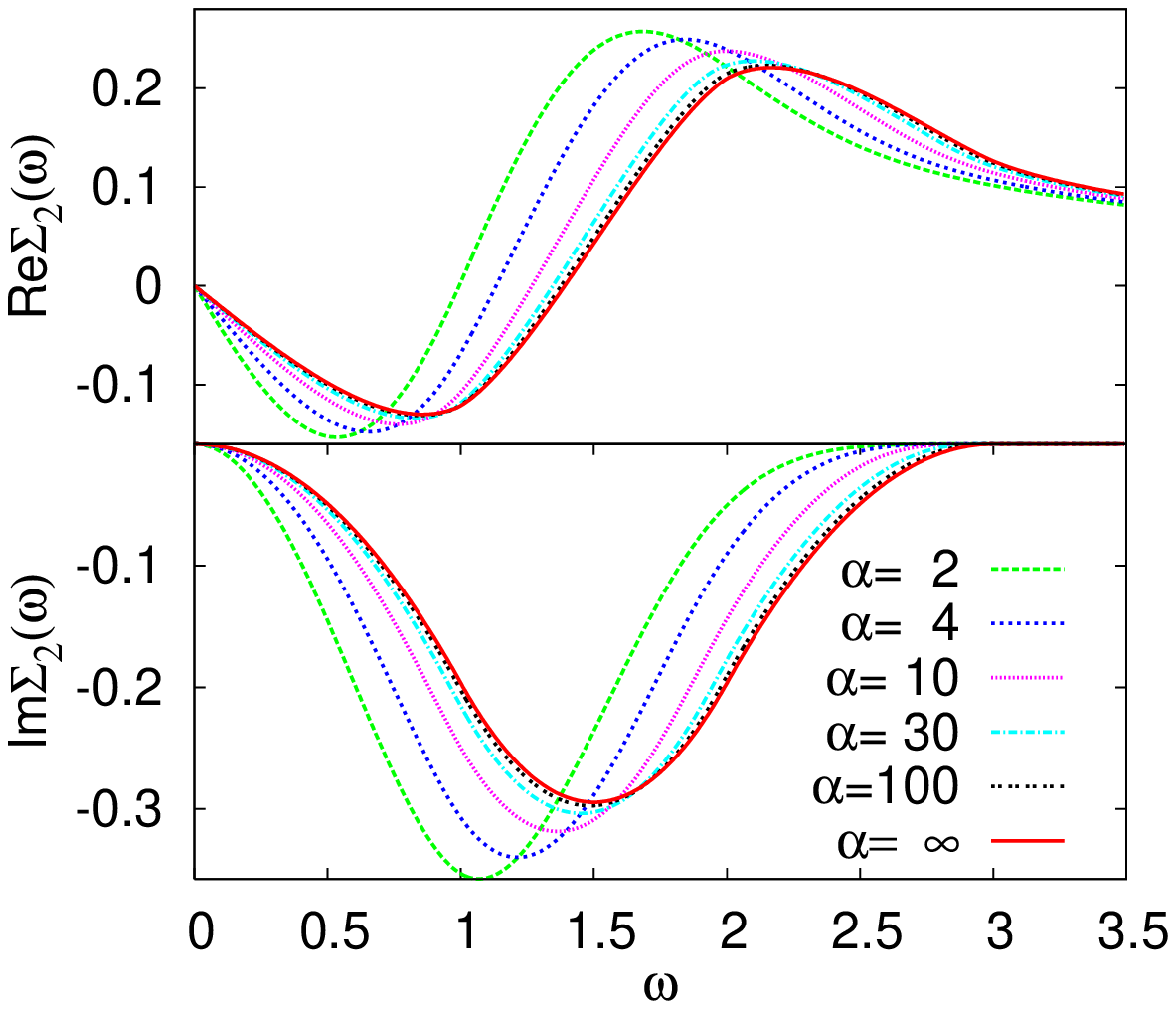}}
    \caption{(a) The model DOS given by Eq.~(\ref{eq:model-dos}). (b)
      The corresponding second order self-energies
      $\Sigma_2(\omega)$.}
    \label{fig:alpha}
  \end{figure}

  The functions $\rho_\alpha(\epsilon)$ are all linear at the
  band-edges but become increasingly steep as $\alpha$ increases
  (Fig.~\ref{fig:alpha}a).  In the limit $\alpha\to\infty$
  $\rho_{\alpha}(\omega)$ approaches the box-shaped DOS
  $\rho_{\infty}(\omega)=\Theta(1-|\omega|)/2$, for which a similar
  calculation as in the last section shows that $\omega_0=\pm 1$.  We
  therefore write $\omega^{(\alpha)}_0=1-\Delta_\alpha/\alpha$ for the
  crossing frequency in the limit $U\to0$ for the DOS
  $\rho_\alpha(\epsilon)$.  As shown in the following,
  $\Delta_\alpha\to\Delta_0$ for $\alpha\to\infty$, i.e., there are
  corrections of the crossing frequency with respect to the band-edge,
  but these corrections become small as the DOS develops a
  discontinuity there.

\begin{figure}
    \centerline{\psfig{clip=true,width=0.85\linewidth,file=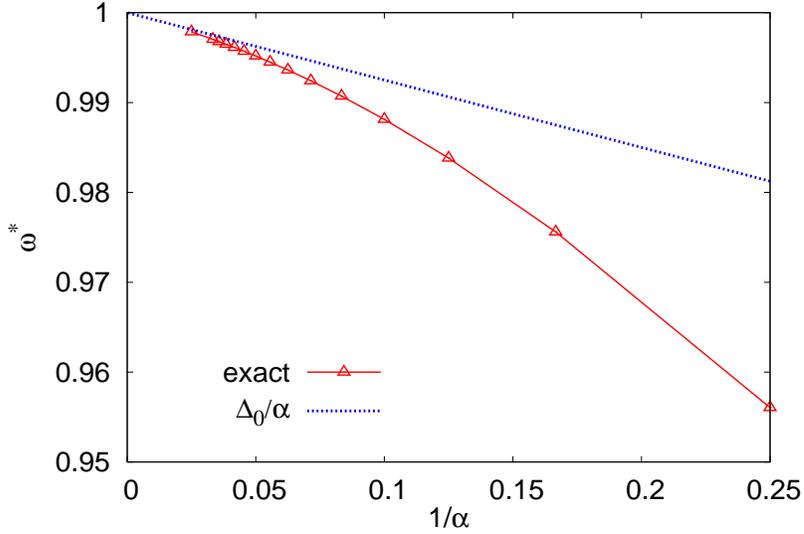}}
    \caption{Location of crossing points in the spectral function
      in the limit $U\to0$ for the
      non-interacting DOS $\rho_\alpha(\epsilon)$
      [Eq.~(\ref{eq:model-dos})]. The triangles mark the numerical
      solution of Eq.~(\ref{eq:a-crossing}), the dotted lines shows
      the lowest order approximation [Eq.~(\ref{eq:lowest-order})].}
    \label{fig:alpha-cross}
  \end{figure}

  We first write Eq.~(\ref{eq:a-crossing3}) as
  \begin{equation}
    \left.
      \frac{(dG_0/dz)'}{(dG_0/dz)''}
    \right|_{z=1-\Delta_\alpha/\alpha+i0}
    =
    -\left.
      \frac{(\Sigma^{(\alpha)}_2)'}{(\Sigma^{(\alpha)}_2)''}
    \right|_{z=1-\Delta_\alpha/\alpha+i0}
    \,.
    \label{eq:eq1}
  \end{equation}
  The second-order contribution to the self-energy is shown in
  Fig.~\ref{fig:alpha}b.  Similar to the last section, the nonanalytic
  behavior of $dG_0/dz$ at $z=1$ determines the behavior of the
  solutions of this equation.  For $\Delta_\alpha \ll \alpha$ and
  $\alpha\to\infty$ the leading contribution to the derivative [see
  Eq.~(\ref{eq:G0})]
  \begin{equation}
    \frac{dG_0}{dz} =
    \frac{\alpha+1}{2\alpha}
    \int_{-1}^{1}dx\,
    \frac{1-x^{\alpha}}{(1-\frac{\Delta_\alpha}{\alpha}+i0-x)^2},
  \end{equation}
  is given by
  \begin{equation}
    \frac{dG_0}{dz} \sim
    -\frac{\alpha}{2}
    \left(
      \frac{1}{\Delta_\alpha}
      +
      \text{Re}\int_{0}^{\infty}dx \frac{e^{-x}}{(x-\Delta_\alpha+i0)^2}
      +
      i\pi e^{-\Delta_\alpha}
    \right).
    \label{eq:dg0dz-expansion}
  \end{equation}
  On the other hand the right-hand side of (\ref{eq:eq1}) is finite
  for $\alpha\to\infty$. We can thus analyze the behavior of the
  solutions for large $\alpha$ by letting $\Delta_\alpha\to0$ and
  $\alpha\to\infty$ in
  $\Sigma_2^{(\alpha)}|_{1-\Delta_\alpha/\alpha}$.  To lowest order we
  then have
  \begin{equation}
    \omega^{(\alpha)}_0 = 1 - \Delta_0/\alpha,
    \label{eq:lowest-order}
  \end{equation}
  where $\Delta_0$ is the a solution of
  \begin{equation}
    e^{\Delta_0}
    \left(
      \frac{1}{\Delta_0}+\text{Re}\int_{0}^{\infty}dx \frac{e^{-x}}{(x-\Delta_0+i0)^2}
    \right)
    =
    \pi \left.
      \frac{(\Sigma^{(\infty)})''}{(\Sigma^{(\infty)})'}
    \right|_{\omega=1}.
    \label{eq:delta0}
  \end{equation}
  Since $\Sigma_2^{(\infty)}(\omega)$ can be calculated directly from
  Eq.~(\ref{eq:sigma2}) and (\ref{eq:a_lambda}) we can solve
  (\ref{eq:delta0}) numerically.  The result, $\Delta_0=0.0750$, is
  compared in Fig.~\ref{fig:alpha-cross} to a direct numerical
  solution of of Eq.~(\ref{eq:sigma2}), (\ref{eq:a_lambda}), and
  (\ref{eq:a-crossing}), which is possible for not too large $\alpha$.
  Fig.~\ref{fig:alpha-cross} clearly demonstrates that in the limit
  $U\to0$ the crossing points are not located at the band-edges if
  there is no van-Hove singularity.

  \section{CONCLUSION}
  \label{sec:concl}

  In this paper we presented an analytical investigation of isosbestic
  points in the local spectral function $A(\omega,U)$ of correlated
  electrons described by the Hubbard model.  The existence of such
  crossing points in the curves of $A(\omega,U)$ vs.\ %
  $\omega$ for different values of the interaction $U$ is due to sum
  rules for $A(\omega,U)$.  For small $U$ the frequency corresponding
  to the isosbestic point may be derived within weak-coupling theory.
  In particular we showed that if the non-interacting DOS has a
  van-Hove singularity at the band-edge, there is in general a
  crossing point at $\omega^* = W+\mathcal{O}(U^2/W)$ for $U\to0$. By
  contrast, if the DOS vanishes linearly at the band-edge the crossing
  frequency is located at $\omega^*(U=0) \ne W$.

  We established that, in general, the crossing frequency $\omega^*$,
  and thus the value of the local spectral function at the crossing
  point, $A(\omega^*)$, sensitively depends on the form of the
  non-interacting DOS $\rho(\epsilon)$ at the band-edges. This is
  quite different from the near-universality of the crossing point
  values of the specific heat at high temperatures.\cite{Chandra99} We
  therefore conclude that the study of isosbectic points in
  \emph{dynamical quantities} can provide valuable information about
  microscopic energy scales of strongly correlated electron systems.

  \section*{ACKNOWLEDGMENTS}

  We thank Ralf Bulla for providing us with his NRG data and Krzysztof
  Byczuk for discussions.  This work was supported in part by the DFG
  Sonderforschungsbereich 484.

\end{document}